\documentclass[reprint,nofootinbib,amsmath,amssymb,showkeys,aps]{revtex4-1}
\usepackage{ulem}
\usepackage{graphicx}
\usepackage{graphicx}
\usepackage{epstopdf}
\usepackage{dcolumn}
\usepackage{bm}
\usepackage{comment}
\usepackage[utf8]{inputenc}
\usepackage{mciteplus}
\usepackage{color}
\usepackage{subfigure}
\usepackage[colorlinks]{hyperref}
\hypersetup{
    linkcolor=blue,
    filecolor=blue,      
    urlcolor=blue,
    citecolor=blue,
}

\begin{document}

 \vspace*{-5mm}
\begin{flushright}
HIP-2018-28/TH
\end{flushright}
\vspace{.3cm}
\thispagestyle{empty}

\title{\boldmath Mimicking features in alternatives to inflation with interacting spectator fields}
\author{Guillem Dom{\`e}nech$^1$}
	\email{domenech@thphys.uni-heidelberg.de}
\author{Javier Rubio$^{1,2}$}
 	\email{javier.rubio@helsinki.fi}
\author{Julius Wons$^1$}
\email{wons@thphys.uni-heidelberg.de}

\affiliation{$^1$ Institut f{\"u}r Theoretische Physik, Ruprecht-Karls-Universit{\"a}t Heidelberg,  Philosophenweg 16, 69120 Heidelberg, Germany }
\affiliation{$^2$ Department of Physics and Helsinki Institute of Physics, \\ 
PL 64, FI-00014 University of Helsinki, Finland}

\begin{abstract}
It has been argued that oscillatory features from spectator fields in the primordial power spectrum could be a probe of alternatives to inflation. In this work, we soften this claim by showing that the frequency and amplitude dependence of the patterns appearing in these scenarios could be mimicked by field interactions during inflation. The degeneracy of the frequency holds for the $n$-point correlation functions, while the degeneracy of the amplitude is broken at the level of non-gaussianities.
\end{abstract}

\keywords{Inflation; Early universe; Primordial features;  Primordial standard clocks; Alternatives to inflation; 	arXiv:1811.08224  }

\maketitle

\section{Introduction}

The stunningly detailed maps provided by numerous Cosmic Microwave Background  observations \cite{Ade:2015lrj} have consolidated inflation \cite{Starobinsky:1979ty,Guth:1980zm,Sato:1980yn,Linde:1983gd} as the leading paradigm for generating the primordial density fluctuations seeding structure formation. In spite of its phenomenological success, some formal questions such as the initial singularity or the trans-Planckian problem remain open for a part of the community.\footnote{These formal aspects are not a problem for the predictions of inflation themselves, as long as inflation is understood as a paradigm and not as a particular model. For an extended  discussion on this issue we refer the reader to  \url{https://blogs.scientificamerican.com/observations/a-cosmic-controversy/}.} This has motivated the quest for alternative scenarios able to generate an initial scale-invariant spectrum of density perturbations without invoking an accelerated expansion of the Universe, but rather an alternative evolution of the scale factor $a$ \cite{Brandenberger:1988aj,Wands:1998yp,Khoury:2001wf,Finelli:2001sr,Gasperini:2002bn,Lehners:2007ac,Battefeld:2014uga,Brandenberger:2016vhg}. 
One of the prototypical examples is a matter contraction era taking place in the very early Universe.  In this scenario, the scale factor $a$ evolves according to a power-law $a\sim t^p$
with $-\infty<t<0$ and $p\approx 2/3$ related to the effective equation--of--state parameter at that epoch, $w_{\rm eff}=2/(3p)-1\approx 0$. Interestingly, due to a duality in the equations of motion for the curvature perturbation \cite{Wands:1998yp}, the spectral tilt of the primordial power spectrum generated by this pre hot big bang evolution \cite{Lucchin:1984yf},
\begin{equation}
    n_s=1+3-\left| 3+\frac{2}{p-1}\right|\approx 1\,,
\end{equation}
turns out to coincide with that generated by an almost de Sitter inflationary expansion with  $w_{\rm eff}\approx-1$ and $\vert p\vert\gg 1$. Several ways of distinguishing inflation from alternatives like this have been proposed in the literature. One possibility is the eventual detection of primordial gravitational waves \cite{Brandenberger:2016vhg} or B-mode polarization \cite{Kamionkowski:2015yta}.
Another option is to look for distinctive features in the primordial spectrum. The main idea behind this approach is that the evolution of a  spectator field during a given 
cosmological era can imprint specific \textit{oscillatory signals} in the density correlation functions. In particular, since the clocking pattern of these probes depends on how the Universe expands (or contracts), there is a chance to determine the evolution of the scale factor by carefully inspecting the amplitude and frequency of the oscillations. 

The evolution of spectator fields during inflation has been extensively studied in the literature \cite{Chen:2009zp,Chen:2011zf,Chen:2011tu,Chen:2014cwa,Chen:2014joa,Chen:2015lza,Chen:2016qce,Brandenberger:2017pjz,Tong:2018tqf,Chen:2018cgg,Raveendran:2018yyh,Achucarro:2010da,Cespedes:2012hu,Gong:2013sma,Arkani-Hamed:2015bza,Chluba:2015bqa,Lee:2016vti,Iyer:2017qzw,Tong:2017iat}. Distinctive oscillatory patterns are found for instance in multi-field inflationary models\footnote{For a review of multi-field inflationary models see for instance \cite{Kaiser:2010yu,Gong:2016qmq} and references therein.} displaying sudden turns in the inflationary trajectory  \cite{Shiu:2011qw,Pi:2012gf,Saito:2012pd,Saito:2013aqa,Gao:2013ota,Konieczka:2014zja,Pi:2017gih,Gundhi:2018wyz} or in single-field scenarios involving steps \cite{Starobinsky:1992ts,Adshead:2011jq} or periodic features in the potential, as in axion monodromy \cite{Silverstein:2008sg,Flauger:2009ab,Flauger:2010ja}. 

Although the idea of using primordial standard clocks as an inflationary test is certainly appealing, the frequency of the oscillations depends implicitly on the detailed structure of the theory. In particular, if the mass of the spectator field gets additional contributions  from other fields, the oscillation frequency would not only depend on the evolution of the scale factor, but also on the additional time-dependence inherited from interactions. To distinguish these field-dependent objects from the standard terminology---which refers to spectator fields with constant masses as \textit{standard clocks}---we will refer to them as (classical) \textit{non-standard clocks}.

In view of the future prospects for detecting oscillatory signals in the Cosmic Microwave Background  \cite{Chen:2016vvw,Bartolo:2017sbu,Gong:2017vve,Palma:2017wxu,Ballardini:2017qwq, Ballardini:2016hpi, Finelli:2016cyd}, it is important to clarify whether those associated with alternative scenarios could be mimicked\footnote{This possibility was pointed out in Ref.~\cite{Chen:2011zf}, although without providing any specific realization.} by non-trivially interacting spectator fields in an inflationary setting. In this paper, we illustrate that this is indeed the case. The manuscript is organized as follows. In Section \ref{sec:bg} we introduce a general framework leading to the appearance of classical non-standard clocks. The associated modifications of the power spectrum are computed in Section \ref{sec:imprints}. The degeneracy of the resulting inflationary signals with those produced in alternatives scenarios is illustrated in Section \ref{sec:mimick}. In Section \ref{sec:higher} we extend our results to higher-order correlation functions. Finally, our conclusions are presented in Section \ref{sec:discussion}. 

\section{Two-field inflationary model \label{sec:bg}}

Let us consider a two-field inflationary model with Lagrangian density
\begin{eqnarray}\label{eq:action}
\frac{\cal L}{\sqrt{-g}} &=&
\frac{M_{\rm P}^2}{2}R 
-\frac{1}{2} \omega^2(\chi)g^{\mu \nu}\partial_{\mu}\phi \partial_{\nu}\phi-V(\phi) \nonumber
\\&-&\frac{1}{2}f^2(\phi)g^{\mu \nu}\partial_{\mu}\chi \partial_{\nu}\chi - \frac{1}{2}m^2(\phi)  \chi^2\,,
\end{eqnarray}
where $M_{\rm P}=1/\sqrt{8\pi G}$ is the reduced Planck mass and  $\omega(\chi)$, $f(\phi)$, $m(\phi)$ and $V(\phi)$ are arbitrary functions of their corresponding arguments. This type of interactions are generically expected to appear upon Weyl rescaling in variable gravity scenarios \cite{Wetterich:2013jsa,Rubio:2017gty} or in models involving non-minimal couplings of the inflaton field to gravity \cite{Kaiser:2010yu,Gong:2016qmq}.\footnote{ Note that, although the naive application of standard effective field theory arguments in Minkowski spacetime would restrict these functions to higher-order polynomials operators suppressed by a given cutoff scale, this is not necessarily the case in the presence of gravity. Indeed, if the aforementioned operators are added in a non-minimally coupled frame, they become exponential-like/dilatonic functions when written in the Einstein-frame.}

For a flat, homogeneous and isotropic Friedmann–Lema{\^i}tre–Robertson–Walker (FLRW) metric, 
\begin{equation}
ds^2=-dt^2+a^2d\mathbf{x}^2\,,
\end{equation}
the equations of motion following from Eq.~\eqref{eq:action} take the form 
\begin{eqnarray}
&& 3M_{\rm P}^2H^2=\frac{1}{2}\omega^2\dot{\phi}^2+ V+\frac{1}{2}f^2\dot{\chi}^2+\frac{1}{2} m^2\chi^2\,, \\
    &&
\frac{1}{a^3}\frac{d}{dt}\left(a^3 \omega^2 \dot\phi \right) 
    +V_{,\phi}+\frac{1}{2}m^2_{,\phi} \chi^2=\frac{1}{2}f^2_{,\phi}\dot \chi^2\,,\label{eq:phi} \\
 &&\frac{1}{a^3}\frac{d}{dt}\left(a^3 f^2 \dot\chi \right) 
    +m^2\chi=\frac{1}{2}\omega^2_{,\chi}\, \dot \phi^2\,,\label{eq:eomchi}
\end{eqnarray}
with $H=\dot a/a$ the Hubble rate and the dots denoting derivatives with respect to the coordinate time $t$. Note that the inclusion of the functions $f$ and $\omega$ changes the effective scale factor experienced by the scalar fields, as can be easily seen by comparing the first terms in Eqs.~\eqref{eq:phi} and \eqref{eq:eomchi} with that of a free scalar field in a FLRW cosmology.

Within this framework, oscillatory patterns in the power spectrum are expected in the presence of sharp turns in field space \cite{Pi:2017gih} or if one of the scalar fields develops a new minimum along its trajectory \cite{Gundhi:2018wyz}.\footnote{We focus here on classical excitation mechanisms, postponing the analysis of quantum ones \cite{Chen:2015lza,Chen:2017ryl} to a future work.} 
In what follows we will focus on the first possibility. In particular, we will assume that the potential $V(\phi)$ renders massive the field $\phi$ during the first stages of inflation while making it light within the observable  Cosmic Microwave Background window. This choice translates into an inflationary dynamic essentially dominated by the $\chi$ field at early times while driven by the $\phi$ field at late times. If sufficiently fast, the transition among these two rolling periods leads to oscillations in the $\chi$ direction that could potentially impact the primordial power spectrum generated by the $\phi$ field. 

Since the dynamics in the presence of a sharp turn is complicated and model dependent, we will focus here on the oscillations of the $\chi$ field during the secondary inflationary stage. To study this $\phi$-dominated period, we will assume the usual slow-roll conditions
\begin{align}\label{eq:approximations}
\epsilon\equiv-\frac{\dot H}{H^2}\ll 1\,, \hspace{10mm} \eta\equiv \frac{\dot \epsilon}{H\epsilon}\ll 1\,.
\end{align}
These conditions, together with the requirement that the $\chi$ field stays subdominant at all times, translate into restrictions on the values and couplings of the $\chi$ field, namely
\begin{align}\label{eq:backreact}
\delta_\omega \ll 1\,,\hspace{15mm} \delta_f f^2\dot\chi^2,\,\,\delta_m m^2\chi^2\ll \omega^2 \dot\phi^2\,,
\end{align}
with the quantities
\begin{align}
\delta_\omega\equiv\frac{d\ln \omega}{dN}\,, \hspace{8mm} \delta_f\equiv\frac{d\ln f}{dN}\,,\hspace{8mm}
\delta_m\equiv\frac{d\ln m}{dN}\,,
\end{align}
measuring the variation of the functions $\omega$,  $f$ and $m$ per Hubble time $dN\equiv Hdt$.  
Note that the latest requirement in Eq.~\eqref{eq:backreact} is not a strong restriction for an oscillating field ($f^2\dot\chi^2\sim m^2\chi^2)$ provided that $\delta_f$ and $\delta_m$ are not excessively large and that the function $\omega^2$ is not too small. 
In order to simplify the analysis, we will assume the latest quantity to be close to the canonical value $\omega^2=1$,
\begin{equation}\label{eq:Deltaw}
\omega^2\equiv1+\Delta \omega^2\,,   \hspace{10mm} \Delta\omega^2\ll 1\,,
\end{equation}
and restrict its potential backreaction effects on the $\chi$-field equation of motion \eqref{eq:eomchi} by requiring
\begin{align}\label{eq:backreact2}
\frac{d\Delta \omega}{d\ln\chi}\ll \frac{m^2}{\epsilon H^2}\frac{\chi^2}{ M_{\mathrm{P}}^2}\,,
\end{align}
with $\epsilon H^2 M_P^2\sim  \dot \phi^2$.

\subsection{Spectator field oscillations}
The oscillations of the $\chi$ field in the $\phi$-field inflationary background are more easily analyzed in terms of a rescaled field $\sigma\equiv fa^{3/2}\chi$, such that the friction terms in Eq.~\eqref{eq:eomchi} are effectively removed. The resulting equation of motion becomes then that of an (undamped) harmonic oscillator,
\begin{align}\label{eq:sigma}
&\ddot\sigma+m_{\rm eff}^2\sigma=0\,,
\end{align}
 with time-dependent mass
\begin{align}\label{eq:effmass}
&m_{\rm eff}^2\equiv \frac{m^2}{f^2}-\frac{\ddot f}{f}-3H\frac{\dot f}{f}-\frac{9}{4}H^2\left(1-\frac{2}{3}\epsilon\right)\,,
\end{align}
where, in agreement with the assumption \eqref{eq:Deltaw}, we have neglected a small $\Delta\omega^2$ contribution.

In usual situations, the time derivatives of the function $f$ are proportional to the instantaneous Hubble rate $H$ and the effective mass $m_{\rm eff}^2$ is dominated by the first term in Eq.~\eqref{eq:effmass}. Assuming this to be the case, we can easily compute the 
solution of Eq.~\eqref{eq:sigma} at the lowest order in the WKB approximation. Defining an expansion parameter 
\begin{align}
\mu^{-1} \equiv \frac{{H}}{m_{\rm eff}}\ll 1\,,
\end{align}
we get 
\begin{align}\label{eq:chi}
\chi=\chi_r\left(\frac{a}{a_r}\right)^{-3/2}\frac{f_r}{f}\sqrt{\frac{m_{\rm eff,r}}{m_{\rm eff}}}\,\sin\left(\Omega+\frac{\theta}{2}\right)+O(\mu^{-1})\,,
\end{align}
with the subindex $r$ referring to the evaluation of the corresponding quantity at the onset of the first oscillation, $\theta$ an (irrelevant) integration constant and 
\begin{align}
\Omega\equiv \int m_{\rm eff} \,dt
\end{align}
an integrated frequency. 

The oscillations in Eq.~\eqref{eq:chi} leave an imprint in the background Hubble rate, which,  in average and at leading order in the expansion parameter $\mu^{-1}$, is only affected by the $\chi$-field mass $m(\phi)$ via the function $\delta_m$,
\begin{align}
\frac{H_{\rm osc}}{H}\approx-\frac{\mu_r}{8}\frac{\chi_r^2}{M_{\rm P}^2}\left(1+\frac{1}{3}\delta_m\right)\frac{H_r}{H}\left(\frac{a}{a_r}\right)^{-3}\sin\left(2\Omega+\theta\right)\,. \label{eq:Hosc}
\end{align}
This result can be easily understood by noticing that the averaged energy density of the $\chi$ field behaves like dust irrespectively of the coupling $f$. Note also that, due to an accidental cancellation, the oscillations of the $\chi$ field for $\delta_m=-3$ enter the background evolution at order ${\cal O}(\mu^{-2})$.

\section{2-point correlation function \label{sec:imprints}}

The presence of the oscillating field $\chi$ modifies the evolution of the (gauge invariant) 
curvature perturbation ${\cal R}$ \cite{Mukhanov:1990me,Kodama:1985bj}. In spite of its extensive use in the literature, we will refrain from using the standard terminology referring to the background contribution \eqref{eq:Hosc} as ``gravitational contribution'' and to that associated with the direct kinetic couplings between $\chi$ and $\phi$ as ``direct contribution''.\footnote{This classification is clearly not gauge invariant, as one can always choose a constant $H$ slicing.} We will rather distinguish two main scenarios according to the behaviour of the scalar interactions during slow-roll. 

\subsection{Slow-roll suppressed interactions}

If $\Delta \omega^2=0$ the modifications in the power spectrum are suppressed by the slow-roll conditions \eqref{eq:backreact}. Among the different contributions induced in the (gauge invariant) curvature perturbation ${\cal R}_k$, the most relevant one is associated with the highest number of time derivatives, since these provide more powers of the large parameter $\mu$. Keeping only this leading contribution, the Mukhanov-Sasaki equation for the mode functions with wavenumber $k$ takes the form of a Mathieu equation, \cite{Mukhanov:1990me,Kodama:1985bj}
\begin{align}\label{eq:MS}
u_k''+\left(k^2-\frac{z''}{z}-\Delta(\tau)\cos\left(2\Omega+\theta\right)\right)u_k=0\,,
\end{align}
with the primes denoting derivatives with respect to the conformal time $\tau$,
\begin{align}
\frac{\Delta(\tau)}{a^2H^2}\equiv
-\frac{\mu_r^4}{4}\frac{\chi_r^2}{\epsilon M_{\rm P}^2}\left(1+\frac{1}{3}\delta_m\right)\frac{H_r}{H}\left(\frac{\mu}{\mu_r}\right)^{3}\left(\frac{a}{a_r}\right)^{-3}\,,
\end{align}
and 
\begin{equation}\label{eq:zdef}
u_k\equiv z{\cal R}_k \,, \hspace{15mm} z\equiv aM_{\rm P}\sqrt{2\epsilon}\,.
\end{equation}
Although studying the resonant band structure of this quasi-periodic equation might be interesting on its own, we will restrict ourselves to the simplest perturbative treatment,\footnote{In this limit, the calculations using the in-in formalism and the equations of motion coincide \cite{Chen:2015dga}.} postponing the non-perturbative analysis to a future publication. In particular, we will expand the Mukhanov-Sasaki mode functions as $u_k=u_{k,0}+u_{k,1}+...$ with  $u_{k,0}$ given by the underlying inflationary model and $u_{k,1}$ a small perturbation satisfying  
\begin{align}\label{eq:uk1}
   u_{k,1}''+\left(k^2-\frac{z''}{z}\right)u_{k,1}=\Delta(\tau)\cos\left(2\Omega+\theta\right)u_{k,0}\,.
\end{align}
The total curvature power spectrum is then given by
\begin{align}
P_{\cal R}=\frac{k^3}{2\pi^2}\lim_{\tau\to0}\left|\frac{u_{k,0}+u_{k,1}}{z}\right|^2=P_{{\cal R},0}+\Delta P_{\cal R}\,,
\end{align}
with
\begin{align}
P_{{\cal R},0}=\frac{k^3}{2\pi^2}\lim_{\tau\to0}\left|\frac{u_{k,0}}{z}\right|^2\,,
\end{align}
and
\begin{align}
\Delta P_{\cal R}=\frac{k^3}{\pi^2}\lim_{\tau\to0}\frac{{\rm Re}\left[u^*_{k,0}u_{k,1}\right]}{z}\,.
\end{align}
For the purposes of this paper it  would be enough to approximate $u_{k,0}$ by the standard de Sitter mode functions
\begin{align}
u_{k,0}=\frac{1}{\sqrt{2k}}\left(1-\frac{i}{k\tau}\right){\rm e}^{-ik\tau}\,,
\end{align}
leaving aside a small correction of the order of slow-roll parameters. 
The solution of Eq.~\eqref{eq:uk1} can be computed using the Green function method and yields \cite{Huang:2016quc}
\begin{equation}
\begin{split}\label{eq:uk1int}
u_{k,1}(\tau)& = iu_{k,0}(\tau)\int^{\tau_f}_{\tau} \mathop{d\xi}\left|u_{k,0}\right|^2\Delta(\xi)\cos\left(2\Omega+\theta\right)
\\
&-
iu^*_{k,0}(\tau)\int_{\tau_r}^\tau \mathop{d\xi}\left(u_{k,0}\right)^2\Delta(\xi)\cos\left(2\Omega+\theta\right)\,,
\end{split}
\end{equation}
where we have assumed homogeneous boundary conditions at the onset of the oscillations. Note, however, that the exact boundary conditions used do not play a central role in the final result. Indeed, the integrals in this expression are dominated by the time at which frequency of the de Sitter mode function $u_{k,0}$ coincides with that of the oscillating field $\chi$. In other words, the resonant behaviour at
\begin{align}\label{eq:res}
K=\Omega'=m_{\rm eff}a\,,
\end{align}
allows to safely evaluate the integral at that time using the saddle-point approximation. 

At this point we have to specify the inflationary model in order to analytically compute the corrections to the primordial power spectrum. For this purpose, and in order to facilitate the comparison with the existing literature, we consider an illustrative (power-law) inflation scenario  with runaway potential $V(\phi)=V_0 {\rm e}^{-\lambda\phi}$  \cite{Lucchin:1984yf}. For this particular choice, the equation of motion of the background field $\phi$ admits a slow-roll 
solution  with 
\begin{align}
a\sim t^p\,, \hspace{10mm}\phi\sim \frac{2}{\lambda}\ln t\,,\hspace{10mm}  p=2/\lambda^2\,.
\end{align}
For simplicity we will also consider constant rate variations of the theory defining functions $f$ and $m$, namely  $\delta_m\,,\delta_f={\rm constant}$. This corresponds to dilatonic-like couplings $m^2\sim {\rm e}^{{\lambda\delta_m}\phi}$ and $f^2\sim {\rm e}^{{\lambda\delta_f}\phi}$, ubiquitously appearing in non-minimally coupled theories when written in the Einstein-frame \cite{Fujii:2003pa,Pi:2017gih}. Under these assumptions, the oscillatory correction to the power spectrum can be written as 
\begin{align}\label{eq:PR}
\frac{\Delta P_{\cal R}}{P_{{\cal R},0}}=&\frac{\sqrt{\pi}\,\mu_r^{5/2}}{4}
\frac{\chi_r^2}{\epsilon M_{\rm P}^2}
\frac{\left(1+\frac{1}{3}\delta_m\right)}{\sqrt{1+\delta_m-\delta_f}}\left(\frac{2K}{k_r}\right)^{\nu_1}\times\nonumber\\&
\sin\left[\frac{2\mu_r}{\gamma}\frac{p^2}{p-1}\left(\frac{2K}{k_r}\right)^\frac{\gamma}{p}+\frac{3\pi}{4}+\theta\right]\,,
\end{align}
where $k_r\equiv2a_rm_{\rm eff,r}$ is the momentum associated with the first resonant mode and we have defined an amplitude tilt
\begin{align}
\nu_1=-3+\frac{5}{2}\frac{\gamma}{p}+2\frac{\delta_m-\delta_f}{1+\delta_m-\delta_f}\,,
\end{align}
with
\begin{align}\label{eq:gamma}
\gamma\equiv\frac{1+p\,\delta_m-p\,\delta_f}{1+\delta_m-\delta_f}\,. 
\end{align}
For $m=f$ (or equivalently $\delta_m=\delta_f$ or $\gamma=1$) the effective mass in Eq.~\eqref{eq:effmass} is (approximately) constant and our results reduce to those in Ref.~\cite{Chen:2011zf}.  Note also that there exists a ``singular'' scenario with $\delta_f-\delta_m=1$ in which the frequency evolves proportional to the conformal time ($\Omega=m_{\rm{eff},r}\tau$) and only one mode resonates. In this case, the saddle-point approximation breaks down and a next-to-leading order computation is required. We will not be interested in this particular setting in what follows.

\subsection{Non slow-roll suppressed interactions}

If $\Delta\omega^2\neq0$ the direct kinetic coupling $\omega^2(\partial\phi)^2$ in  Eq.~\eqref{eq:action} can easily dominate over the other interaction terms,  since, contrary to them, it is not suppressed by the slow-roll conditions \eqref{eq:backreact}. 

Since we want the backreaction of the $\chi$ field on the background dynamics to stay small during the whole observable window, we will  assume the conditions \eqref{eq:Deltaw} and \eqref{eq:backreact2} to hold and consider a simple perturbation  \cite{Saito:2012pd} 
\begin{align}
\Delta\omega^2=\frac{\chi}{\Lambda}\,,
\end{align}
with $\Lambda$ a given energy scale. When this coupling is taken into account the variable $z$ in Eq.~\eqref{eq:zdef} is effectively replaced by $z\,\omega$, leading to an additional  mass term proportional to $\chi''$ in the Mukhanov-Sasaki equation \eqref{eq:MS}. Taking into account Eq.~\eqref{eq:chi} we obtain
\begin{align}\label{eq:MS2}
u_k''+\Big(k^2-\frac{z''}{z}-\Delta_T(\tau)\Big)u_k=0\,,
\end{align}
with 
\begin{equation}
\Delta_T(\tau)=\Delta(\tau)\cos\left(2\Omega+\theta\right)\nonumber+\Delta_\omega(\tau)\sin\left(\Omega+\theta/2\right)\,,    
\end{equation}
and
\begin{align}
\frac{\Delta_\omega(\tau)}{a^2H^2}\equiv -\mu_r^{5/2}\frac{\chi_r}{\Lambda}\frac{f_r}{f}\left(\frac{H_r}{H}
\right)^{1/2}\left(\frac{\mu}{\mu_{r}}
\right)^{3/2}\left(\frac{a}{a_{r}}\right)^{-3/2}\,.
\end{align}
Performing a similar computation to that in the $\Delta \omega^2=0$, we obtain an oscillatory correction to the power spectrum displaying the same frequency dependence\footnote{Strictly speaking, the frequency is halved with respect to the previous case since the corrections depend now on $\chi$ instead of $\chi^2$.} 
but a different amplitude tilt, namely 
\begin{align}\label{eq:nu1}
\nu_1=-\frac{3}{2}+\frac{1}{2}\frac{\gamma}{p}+\frac{\delta_m -2\delta_f}{1+\delta_m-\delta_f}\,.
\end{align}
This scenario is particularly interesting since for $m=f^2$ (or equivalently $\delta_m=2\delta_f$) the signal mimics that of a model with power-law index $\bar p \equiv p/\gamma$. This  correspondence is expected since the equation of motion for the $\chi$ field  effectively reproduces in this case that of a FLRW universe with scale factor $\bar a =f a$ \cite{Li:2014qwa,Domenech:2015qoa}. 
Note indeed that the frequency of the oscillations in the curvature perturbation is determined by the ratio of two time scales: the effective mass $m_{\rm eff}$ and the expansion rate $H$. For a constant mass, the change in $\mu=m_{\rm eff}/H$  is solely due to the Hubble rate $H$ and therefore the frequency probes directly the expansion rate of the Universe. When the mass gets a time dependence we see a combination of both. The time dependence of the mass can be, however, absorbed into a new redefined Hubble rate $\bar H$, namely
\begin{equation}
    \mu=\frac{m_{\rm eff,r}}{\bar H}\qquad{\rm where}\qquad \bar H\equiv H \frac{m_r}{m}\frac{f}{f_r}\,.
\end{equation}
When $m=f^2$ the quantity $\bar H$ coincides with the expansion rate of a universe  with scale factor $\bar a = f a$. That is the reason why a time dependent spectator field mass is able to mimic the frequency of the oscillating features above. Note that this intuitive argument is not linked to matter domination. In particular, an exponential/dilatonic choice of the theory defining functions $f(\phi)$ and $m(\phi)$ is able to mimic any oscillatory feature generated by standard spectator fields in cosmologies with power-law scale factor evolutions, as, for instance, ekpyrotic scenarios with  $0<p\ll 1$.

\begin{figure}
    \centering
    \includegraphics[width=0.5\textwidth]{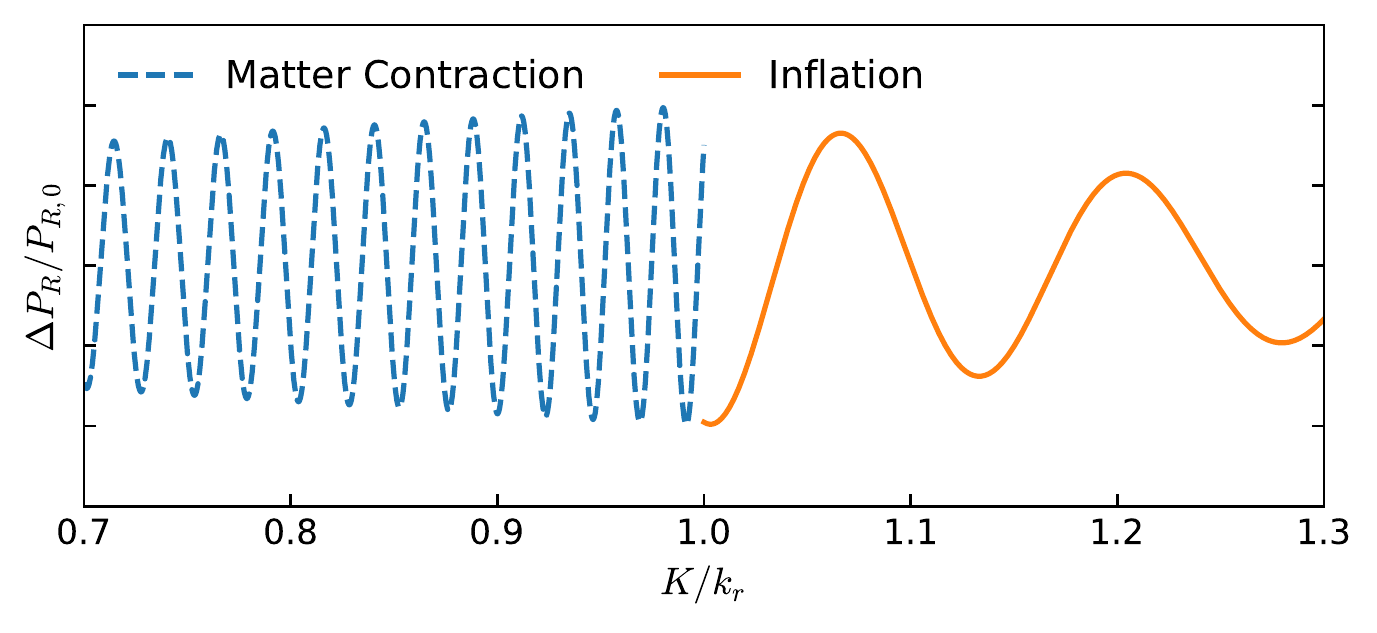}
    \caption{Illustration of the generic standard clock signals (i.e. constant $m$, $f$ and $w$) appearing in matter contraction (blue line) and inflationary scenarios (orange line). We have fixed the value of $k_r$ and, thus, the signal of the matter contraction evolves towards smaller $K$  values since the frequency of the oscillations ($K=m_{\rm eff}a$) decreases in a contracting phase.}
    \label{fig:ConInv}
\end{figure}

\section{Mimicking alternative scenarios}\label{sec:mimick}

The typical behaviours of the oscillatory signals generated by a free spectator field during a matter contraction era and during an inflationary stage are illustrated in  Fig.~\ref{fig:ConInv}. Let us discuss how this picture is modified in the presence of non-trivial interactions.  To this end, note that by taking
\begin{equation}
    \delta_m-\delta_f=\frac{1-\bar p/p}{\bar p -1}\,.
\end{equation}
one can always adjust the scale dependence of the frequency to that of a universe with a power-law evolution $a\sim t^{\bar p}$.
\begin{figure}
    \centering
    \includegraphics[width=0.49\textwidth]{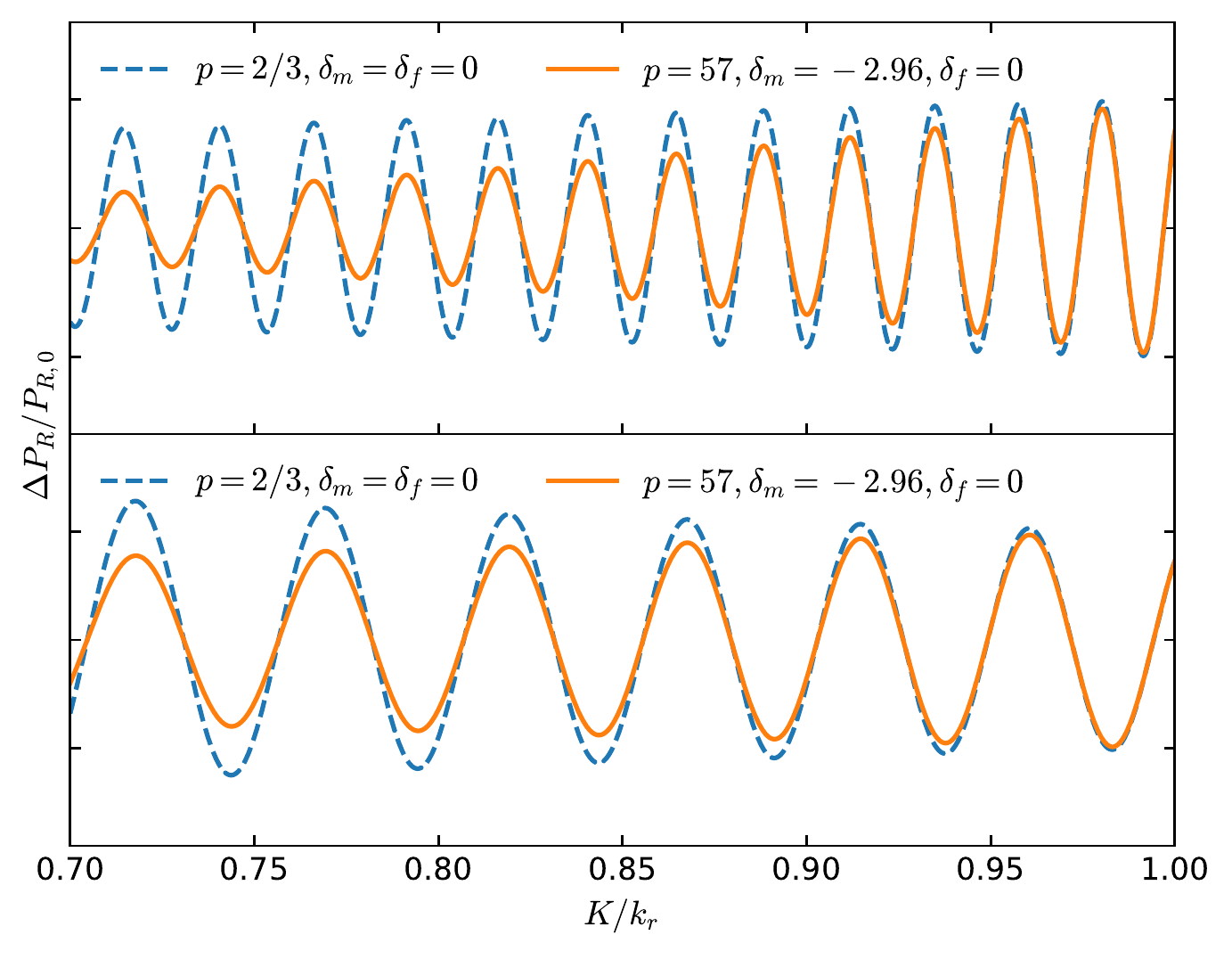}
    \caption{Comparison of the corrections to the power spectrum associated with a standard clock signal during matter contraction (blue line) and a non-standard one during inflation and  with the \textit{same frequency} (orange line). The cases $\Delta \omega^2=0$ and $\Delta\omega^2\neq 0$ are displayed in the upper and lower panel, respectively. For the matter contraction era we choose $\mu_r=25$  and $p=2/3$, while for inflation we take $\mu_r=50$, $p=57$ and $\delta_m=-2.96$.  For an easier comparison, we left the amplitude in arbitrary units and fixed  to be the same at $K=k_r$.
    }
    \label{fig:nonStandard}
\end{figure}
\begin{figure}
    \centering
    \includegraphics[width=0.5\textwidth]{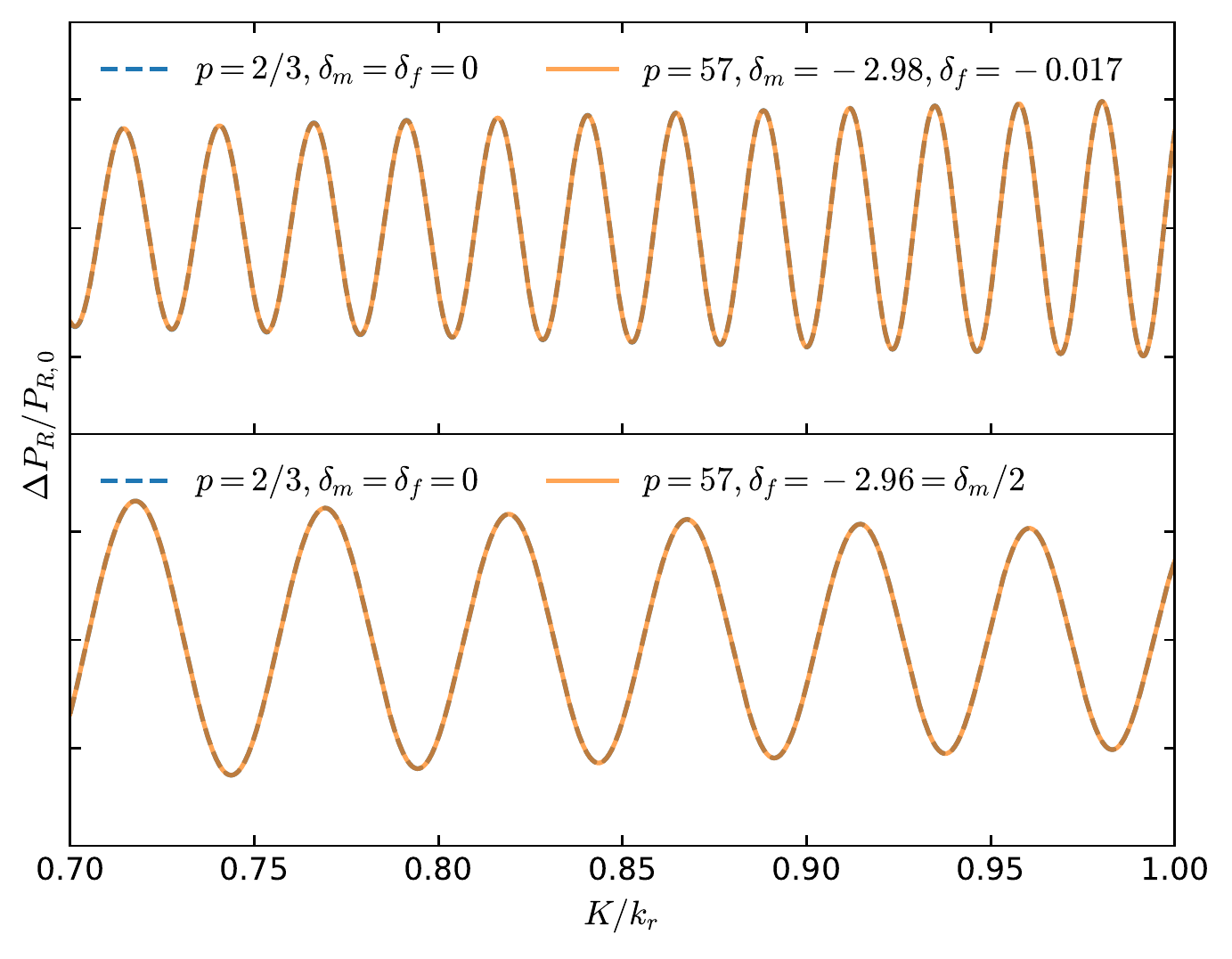}
    \caption{Comparison of the corrections to the power spectrum associated with a standard clock signal during matter contraction (blue line) and a non-standard one with the \textit{same frequency and amplitude} during inflation (orange line). In the top panel we compare the signals coming from a $\Delta\omega^2=0$ scenario in matter contraction and from a $\Delta\omega^2\neq 0$ scenario in inflation. We chose $\mu_r=25$  and $p=2/3$ for the matter contraction case and $\mu_r=50$, $p=57$, $\delta_m=-2.98$ and $\delta_f=-0.017$ for the inflationary case. In the lower panel, we compare the signal coming from two scenarios with $\Delta\omega^2\neq 0$. We choose $\mu_r=25$  and $p=2/3$ for the matter contraction case and $\mu_r=50$, $p=57$ and $\delta_f=-2.96=\delta_m/2$ for the inflationary one. In order to facilitate the comparison, we left again the amplitude in arbitrary units and fixed to be the same at $K=k_r$.}
    \label{fig:Deg}
\end{figure}
For a fixed wavenumber $k_r$, the frequency rescales as well, mimicking that of a field with effective mass 
\begin{align}
    \bar \mu_r\equiv\mu_r\left|\frac{p}{\bar p}\,\frac{\bar p -1}{p-1}\right|\,.
\end{align}
The simultaneous choice of $\delta_m$ and $\delta_f$ allows therefore to fix the frequency and tilt of the oscillations in order to imitate alternative signals. However, if $\Delta \omega^2=0$ the scale dependence of the amplitude breaks the degeneracy since both $\nu_1$ and $\bar{p}=p/\gamma$ depend only on the combination $\delta_m-\delta_f$ (cf. Fig.~\ref{fig:nonStandard}). On the other hand, if the term proportional to $\Delta\omega^2$ dominates, the dependence of tilt on $\delta_m$ and $\delta_f$ changes and we can choose $\delta_m=2\delta_f$ such that the amplitude of the oscillations mimics also that a universe with power-law index $\bar p$, i.e. $\nu_1=-3/2+1/(2\bar p)$ (cf. Fig~\ref{fig:Deg}). 

Incidentally, we have found a degeneracy between the gravitational signal from a matter contraction and a direct signal from inflation with $p\approx 57$, $\delta_m\approx-2.98$ and $\delta_f\approx-0.017$. As we will see in the next section, this accidental degeneracy does not hold  when higher-order correlations are taken into account.
Note also that if $\delta_f>\delta_m$ the effective mass of the field $\chi$ decreases with time, such that it might eventually dominate the background evolution. We assume that such regime is never reached.

\begin{figure}
    \centering
    \includegraphics[width=0.5\textwidth]{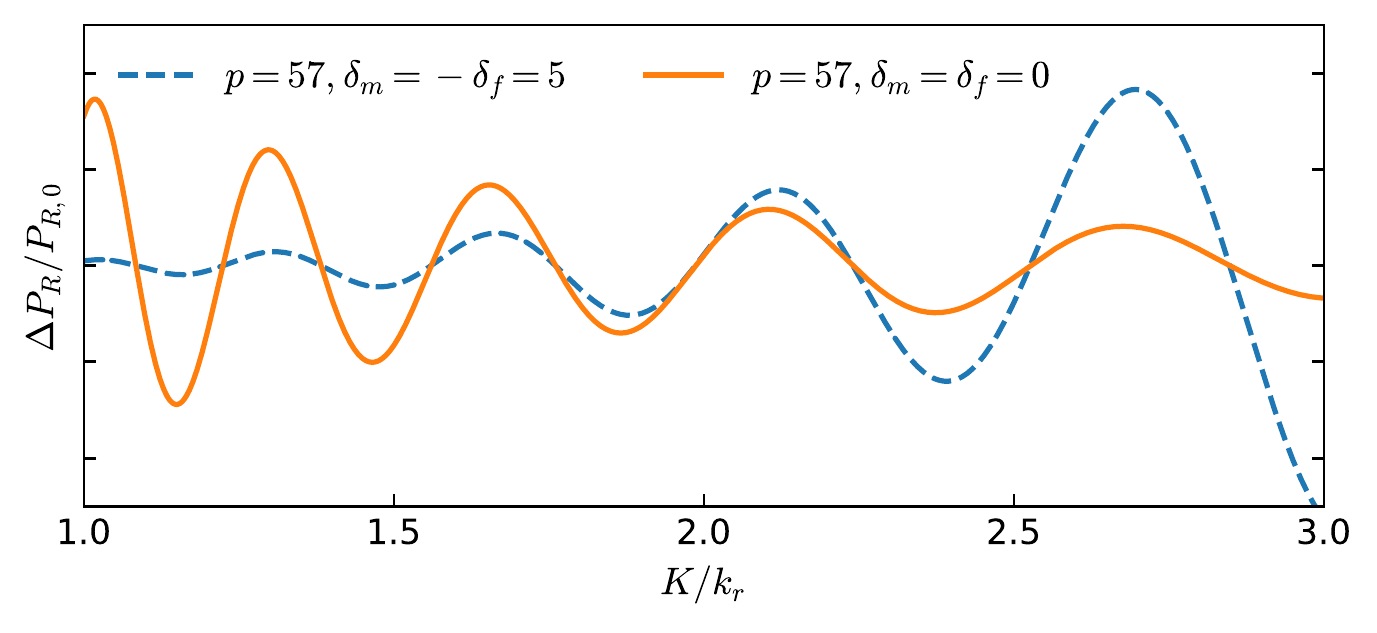}
    \caption{Comparison of particular signal coming from a non-standard clock  with $\Delta\omega^2\neq 0$ (blue line) and a standard clock one (orange line). For both signals we choose $\mu_r=25$ and $p=57$. For the non-standard clock we  take $\delta_m=-\delta_f=5$.}
    \label{fig:Grow}
\end{figure}

Before ending this section, we would like to point out a particular scenario, only present for non-standard clocks like the ones under consideration. That would be an oscillatory signal with $\gamma=1$ (i.e. with a frequency that could be interpreted as that of an standard clock during inflation) but with an increasing amplitude. This will happen whenever $\delta_f<(1-3p)/2p$ (cf. Fig.~\eqref{fig:Grow}).

\section{Higher-order correlations}\label{sec:higher}

The above discussion illustrates how the combination of sudden turns and time-dependent masses can easily produce varying-frequency oscillatory patterns in the power spectrum mimicking the scale dependence of alternative inflationary scenarios. Given this result,  it is important to check if higher-order correlation functions could allow to break the degeneracy among these scenarios. 

The overall scale dependence of the three-point correlation function can be studied by looking at the equilateral configuration. As before, we consider two cases according to the behaviour of the scalar interactions during slow-roll. For $\Delta \omega^2=0$, the leading contribution to the bispectrum  reads \cite{Maldacena:2002vr,Koyama:2010xj,Wang:2013eqj}
\begin{align}
{\cal L}_{3,I}\propto a^3\,M_{\rm P}^2\epsilon\,\dot\eta\, {\cal R}^2\dot{\cal R}\,,
\end{align}
with $\epsilon$ and $\eta$ the first and second slow-roll parameters.
Using the in-in formalism, we find that, up to a shape function, the associated non-gaussianity is given by\footnote{Note that if $m=f$ the effective mass is constant and our results coincide with those in Refs.~\cite{Chen:2011zf} and \cite{Chen:2015lza}.}
\begin{align}\label{eq:NG}
f^{\rm osc}_{NL}\approx&\frac{5\sqrt{\pi}\mu_r^{9/2}}{48 } \frac{\chi_r^2}{\epsilon M_{\rm P}^2}
\frac{\left(1+\frac{\delta_m}{3}\right)}{\sqrt{1+\delta_m-\delta_f}}\left(\frac{K}{k_r}\right)^{\nu_2}\times\nonumber\\&
\sin\left[\frac{2\mu_r}{\gamma}\frac{p^2}{p-1}\left(\frac{K}{k_r}\right)^\frac{\gamma}{p}+\frac{3\pi}{4}+\theta\right]\,,
\end{align}
with $K\equiv k_1+k_2+k_3$ and
\begin{align}
\nu_2=-3+\frac{7}{2}\frac{\gamma}{p}+3\frac{\delta_m-\delta_f}{1+\delta_m-\delta_f}\,. 
\end{align}
On the other hand, if $\Delta \omega^2\neq 0$ the leading contribution to the three-point correlation function is rather given by \cite{Saito:2013aqa}
\begin{align}
{\cal L}_{3,I}\propto a^3\,M_{\rm P}^2\frac{\chi}{\Lambda}\epsilon^2 {\cal R}\dot{\cal R}^2\,.
\end{align}
In this case the scale dependence of the amplitude is modified to
\begin{align}
\nu_2=-\frac{3}{2}+\frac{1}{2}\frac{\gamma}{p}+\frac{2\delta_m-3\delta_f}{1+\delta_m-\delta_f}\,.
\end{align}
We see therefore that even in the case $m=f^2$ there is a correction to the scale dependence of the amplitude of the non-gaussianities as compared to that in an scenario with power-law index $\bar p$. This implies that, at least in the simple case we studied, the degeneracy would be broken by an eventual measurement of the scale dependence of non-gaussinities. 
 
Independently of their origin, the generated non-gaussianities display the same frequency dependence as the power spectrum. This result is due to the existence of a single oscillation frequency and can be easily extended to higher-order correlation functions. We conclude therefore that, if the existence of oscillatory features in the power spectrum would be eventually established, one would need the full information of the amplitudes and the frequency in order to probe the evolution of the scale factor. 

\section{Conclusions and discussion\label{sec:discussion}}

The frequency and momentum dependence of potential oscillatory features in the primordial density fluctuations has been advocated as a way to distinguish the inflationary paradigm from alternatives scenarios such as bounces or matter contraction eras \cite{Chen:2009zp,Chen:2011zf,Chen:2011tu,Chen:2014cwa,Chen:2014joa,Chen:2015lza,Chen:2016qce,Brandenberger:2017pjz,Tong:2018tqf,Chen:2018cgg,Raveendran:2018yyh}. These features are typically imprinted by the oscillations of a spectator field, provided that this is excited at some point in the primordial Universe. 

In this work, we have shown that if the mass of a spectator field receives an explicit time dependence  during inflation---due for instance to non-trivial interaction with other fields---the oscillatory signal in the power spectrum and other higher-order correlation functions can easily mimick the one appearing in alternatives scenarios to inflation, both in its frequency and momentum dependence.
To illustrate this result we considered two particular examples, one with no dominant kinetic mixing among the spectator field and the inflaton kinetic term and one involving a relevant kinetic coupling $\chi(\partial\phi)^2$. Using a WKB approximation we showed that the oscillations in the power spectrum take the generic form
\begin{align}\label{eq:sum1}
\frac{\Delta P_{\cal R}}{P_{\cal R}}\propto \left(\frac{K}{k_r}\right)^{\nu_1}\sin\left[C\left(\frac{K}{k_r}\right)^\frac{\gamma}{p}+\theta\right]
\end{align}
with $C$ and $\theta$ constants, $p$ the power-law index of the scale factor and $\gamma$ a parameter related to the time-dependent spectator field mass and its kinetic term normalization. Interestingly, only the combination $p/\gamma$ is probed by the frequency, but not just the power-law index $p$. The scale dependence of the amplitude in Eq.~\eqref{eq:sum1} is given by
\begin{align}
\nu_1=
\begin{cases}\vspace{3mm}
-3+\frac{5}{2}\frac{\gamma}{p}+2\frac{\delta_m-\delta_f}{1+\delta_m-\delta_f}\hspace{5mm} 
\textrm{if} \hspace{5mm} \Delta \omega^2 =0\,,\\
-\frac{3}{2}+\frac{1}{2}\frac{\gamma}{p}+\frac{\delta_m-2\delta_f}{1+\delta_m-\delta_f}
\hspace{6mm} \textrm{if} \hspace{5mm} \Delta \omega^2 \neq 0\,,
\end{cases}
\end{align}
with $\delta_m$ and $\delta_f$ free parameters of the model related to the change rate of the mass and the kinetic coefficient of the $\chi$ field, respectively. In the former case ($ \Delta \omega^2 =0$), one can choose the interaction among fields in such a way that the frequency matches that of an alternative matter contraction scenario, but the scale dependence of the amplitude breaks the degeneracy. In the second case  ($\Delta \omega^2 \neq0$), the amplitude of the power spectrum can also follow that of the alternative matter contraction scenario for a particular choice of the interactions. Regarding the frequency, this result holds even when one considers higher order correlations functions, since the single oscillation frequency of the spectator field gets imprinted in the same way in all of them. However, the degeneracy is broken by the scale dependence of the amplitude of higher-order correlations functions. We illustrated this result by explicitly computing the bispectrum in the equilateral configuration,  which takes the schematic form
\begin{align}
f_{\rm NL}\propto\left(\frac{K}{k_r}\right)^{\nu_2}\sin\left[C\left(\frac{K}{k_r}\right)^\frac{\gamma}{p}+\theta\right]\,,
\end{align}
with
\begin{align}
\nu_2=
\begin{cases}\vspace{3mm}
-3+\frac{7}{2}\frac{\gamma}{p}+3\frac{\delta_m-\delta_f}{1+\delta_m-\delta_f} \hspace{5mm} 
\textrm{if} \hspace{5mm} \Delta \omega^2 =0\,,\\
-\frac{3}{2}+\frac{1}{2}\frac{\gamma}{p}+\frac{2\delta_m-3\delta_f}{1+\delta_m-\delta_f}\hspace{7mm} \textrm{if} \hspace{5mm} \Delta \omega^2 \neq 0\,.
\end{cases}
\end{align}
The mimicking mechanism proposed in this paper does not require special fine-tunings or large number of parameters. Although we focused for concreteness on matter contraction scenarios leading to an almost scale invariant primordial power spectrum, our results can be easily extended to other alternative scenarios by properly choosing the form of the interactions.  Our findings soften the claim that oscillatory features from spectators fields during inflation can probe inflation and its alternatives without a \textit{full} knowledge of the $n$-point correlation functions. To do so, one would need the complete information of the frequency and amplitudes of, at least, the $2$- and $3$-point correlation functions.

\section*{Acknowledgments}
G.D. would like to thank R.~Namba, M.~Sasaki and S.~Zhou for useful discussions. G.D. would also like to thank X.~Chen for suggesting the term ``non-standard clock''.
G.D was partially supported by DFG Collaborative Research center SFB 1225 (ISOQUANT). 
J.R. acknowledges support from the DFG through the project TRR33 ``The Dark
Universe'' during the first stages of this work.

\bibliographystyle{apsrev}

\bibliography{biblio}

\end{document}